\documentclass[aps,reprint,superscriptaddress,notitlepage]{revtex4-2}
\usepackage[utf8]{inputenc}
\usepackage{xprintlen}
\usepackage{amsmath}
\usepackage{physics}
\usepackage[version=4]{mhchem}
\usepackage{url}
\usepackage{amsfonts}
\usepackage{amssymb}
\usepackage{tabularx}
\usepackage{mathtools}
\usepackage{subfigure}
\usepackage{booktabs}
\usepackage{makeidx}
\usepackage{graphicx}
\usepackage{lmodern}
\usepackage{siunitx}
\usepackage{upgreek}
\usepackage{adjustbox}
\usepackage{blindtext}
\usepackage{siunitx}
\usepackage[section]{placeins}
\usepackage{tikz}
\usepackage{tkz-euclide}
\usetikzlibrary{calc}
\usetikzlibrary{optics}
\usepackage{float}
\usepackage{braket}
\usepackage{appendix}
\usepackage{xr}
\begin{document}
\title{Investigation and comparison of measurement schemes in the low frequency biosensing regime using solid-state defect centers}
\author{Andreas F.L. Poulsen}
\affiliation{Center for Macroscopic Quantum States (bigQ), Department of Physics, Technical University of Denmark, 2800 Kongens Lyngby, Denmark}%
\author{James L. Webb}
\email{jaluwe@fysik.dtu.dk}
\affiliation{Center for Macroscopic Quantum States (bigQ), Department of Physics, Technical University of Denmark, 2800 Kongens Lyngby, Denmark}%
\author{Kirstine Berg-S{\o}rensen}
\affiliation{Department of Health Technology, Technical University of Denmark, 2800 Kongens Lyngby, Denmark}%
\author{Ulrik Lund Andersen}
\email{ulrik.andersen@fysik.dtu.dk}
\affiliation{Center for Macroscopic Quantum States (bigQ), Department of Physics, Technical University of Denmark, 2800 Kongens Lyngby, Denmark}%
\author{Alexander Huck}
\email{alexander.huck@fysik.dtu.dk}
\affiliation{Center for Macroscopic Quantum States (bigQ), Department of Physics, Technical University of Denmark, 2800 Kongens Lyngby, Denmark}%

\begin{abstract}
Ensembles of solid state defects in diamond make promising quantum sensors with high sensitivity and spatiotemporal resolution. The inhomogeneous broadening and drive amplitude variations across such ensembles have differing impacts on the sensitivity depending on the sensing scheme used, adding to the challenge of choosing the optimal sensing scheme for a particular sensing regime. In this work, we numerically investigate and compare the predicted sensitivity of schemes based on continuous-wave (CW) optically detected magnetic resonance (ODMR) spectroscopy, $\pi$-pulse ODMR and Ramsey interferometry for sensing using nitrogen-vacancy centers in the low-frequency ($<10$ kHz) range typical for signals from biological sources. We show that inhomogeneous broadening has the strongest impact on the sensitivity of Ramsey interferometry, and drive amplitude variations least impact the sensitivity of CW ODMR, with all methods constrained by the Rabi frequency. Based on our results, we can identify three different regions of interest. For inhomogeneous broadening less than 0.3 MHz, typical of diamonds used in state of the art sensing experiments, Ramsey interferometry yields the highest sensitivity. In the regime where inhomogeneous broadening is greater than 0.3 MHz, such as for standard optical grade diamonds or in minaturized integrated devices, drive amplitude variations determine the optimal protocol to use. For low to medium drive amplitude variations, the highest sensitivity is reached using $\pi$-pulse ODMR. For high drive amplitude variations, relevant for widefield microscopic imaging, CW ODMR can yield the best sensing performance. 
\end{abstract}
\maketitle

\section{Introduction}
\begin{figure}
    \centering
    \includegraphics[width=0.45\textwidth]{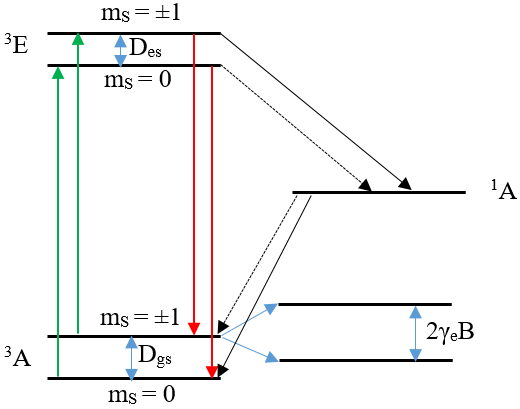}
    \caption{Simplified energy-level diagram of the negatively charged nitrogen-vacancy center (NV). The diagram illustrates the spin-conserving excitation from ground state triplet $^3\mathrm{A}$ to excited state triplet $^3\mathrm{E}$ caused by green illumination, the radiative (red fluorescence) decay back to the ground state triplet and the non-radiative decay paths to and from the metastable singlet state $^1\mathrm{A}$. Dashed arrows indicate relatively weaker decay rates. The diagram also exemplifies the splitting induced by the Zeeman effect when a magnetic field $B$ is applied. We show the zero-field splitting in the ground state, $D_{gs}=2.87$ GHz, and excited state, $D_{es}=1.42$ GHz, and the splitting of the $m_S=\pm1$ states due to a magnetic field $B$ given by the gyromagnetic ratio $\gamma_e=28$ MHz/mT.}
    \label{fig:NVDiagram}
\end{figure}

Defect centers in diamond are promising candidates for applications in quantum sensing due to high attainable sensitivity, their atomic scale dimensions, the chemical stability of diamond and compatibility with biological samples~\cite{Schroder2016, Aharonovich2016, Kucsko2013, Schirhagl2014}. In particular, the negatively charged nitrogen-vacancy center (NV) in diamond possesses several properties advantageous for sensing including long coherence times of the associated electron spin even at room temperature, optical initialization and readout of the spin state~\cite{Loubser1978, Gruber1997} and the ability to coherently manipulate the spin with resonant microwaves (MW)~\cite{Jelezko2004}. The level structure and spin-state dependent transitions of the NV center, illustrated in Fig.~\ref{fig:NVDiagram}, render the system sensitive to temperature~\cite{Neumann2013, Delord2017}, pressure~\cite{Doherty2014}, electric fields~\cite{Dolde2011, Wang2012} and magnetic fields~\cite{Jelezko2004}, but it has received the most focus for its potential as a magnetometer~\cite{Jelezko2004, Taylor2008, Farfurnik2016, Farfurnik2018, Genov2020}. 

NV center based sensing most commonly involves applying a static magnetic field to lift the degeneracy of the $m_S=\pm 1$ spin states and addressing either the $m_S=0 \leftrightarrow m_S =1$ or $m_S=0 \leftrightarrow m_S =-1$ transition as an effective two-level system~\cite{Jelezko2006}. This two-level system is then either driven continuously with constant amplitude microwave and pump laser illumination or in a pulsed manner according to the sensing scheme applied for the specific task. Due to the spin-state dependent fluorescence properties of the NV center, when microwaves are supplied at a resonance frequency matching the triplet ground state spin transition, fluorescence emission is reduced due to nonradiative and infrared decay via a singlet shelving state. Sensing can be performed by recording this environmental-dependent change in fluorescence while probing the spin resonance. This can be practically achieved by many different sensing schemes, including continuous-wave optically detected magnetic resonance spectroscopy (CW ODMR)~\cite{Zhang2020}, $\pi$-pulse ODMR~\cite{Dreau2011, Wolf2015}, Ramsey interferometry~\cite{Rondin2014}, Hahn Echo~\cite{Hong2013} and many others\cite{Jelezko2006, Barry2020, Fescenko2020}. 

Choosing the right scheme for the specific sensing task is essential to achieving the best possible sensitivity. To maximize bulk sensitivity, it is necessary to use a large ensemble of NV centers as in the shot-noise limited regime~\cite{Taylor2008} the sensitivity is proportional to $\sqrt{N}$, where $N$ is the number of NV centers. It is also necessary to use a large ensemble for other applications, in particular wide field of view microscopy with high spatial resolution. The use of an NV center ensemble introduces challenges of its own. Variations in the local environment of NV centers, e.g. related to varying crystal strain across the ensemble, lead to inhomogeneous broadening (IHB) of their spin triplet transition frequencies, making it challenging for a single-frequency MW drive to be on-resonance with the entire ensemble. Depending on the physical extent of the ensemble and the used MW antenna geometry, it is also possible for the MW drive amplitude to vary across the ensemble. Both of these effects can have a negative impact on the performance of sensing schemes. The impact of sample or drive scheme inhomogeneity can also vary between different sensing schemes, which further complicates the choice of the optimal sensing scheme for a given task.

In this article, we numerically investigate and compare the sensitivity of different sensing schemes for different levels of inhomogeneous broadening and drive amplitude variations. We note that for high frequency AC field sensing, such as sensing nuclear spins using single NV centres~\cite{Laraoui2010, Laraoui2011}, spin-echo techniques offer by far the best sensitivity achievable. In this work, we instead focus on the DC to low frequency ($<10$ kHz) sensing regime. This frequency range is of particular importance for biosensing, including sensing of action potentials on the millisecond timescale~\cite{Barry2016} and seconds to hours temperature sensing using nanodiamonds within cells~\cite{Kucsko2013, Fujiwara2020}. This is a primary application for NV centers due to the high degree of biocompatibility of diamond. We consider three sensing schemes: Ramsey interferometry, $\pi$-pulse ODMR and CW ODMR. These are the most commonly used schemes for this frequency range using the NV center platform~\cite{Rondin2014}. We theoretically investigate and compare the predicted relative sensitivities for these three schemes in varying sensing regimes and discuss how these simulations can relate to potential applications, including widefield microscopy and microfabricated sensors. 

\section{Methods}
\begin{figure*}
    \centering
    \includegraphics[width=1.0\textwidth]{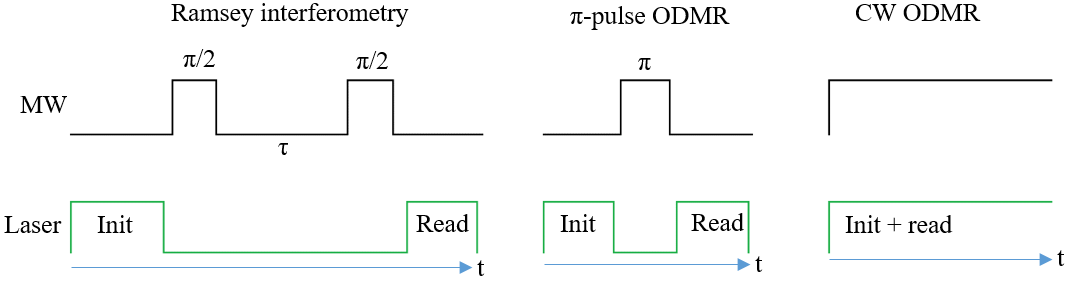}
    \caption{Simple sketch illustrating the application of each of the three considered sensing schemes. Ramsey interferometry and $\pi$-pulse ODMR are pulsed sequences while CW ODMR involves constant application of both laser power and microwaves.}
    \label{fig:sequence}
\end{figure*}

We model an ensemble of $N$ nitrogen-vacancy centers as the sum of independent single NV, denoted $i$, following the level structure in Fig.~\ref{fig:NVDiagram} and following the state dynamics of $N$ independent quantum systems. This assumption is valid due to the relatively low density of NV centers in a typical sensing sample (up to parts per million range), meaning the sum total of interaction between NV centers in the ensemble is low. We assume zero noise, such that our sensitivity is only dependent on the response of the ensemble, excluding difficult to quantify noise sources such as time varying background magnetic fields from laboratory equipment or laser technical noise. The time varying application of microwave and laser fields to the ensemble for each of the three sensing schemes considered in this article are illustrated in Fig.~\ref{fig:sequence}.The exact theoretical approach of the simulations differs between the three considered sensing schemes, but we include the effects of inhomogeneous broadening and MW drive amplitude variations across the ensemble in the same way in all three protocols.

We simulate inhomogeneous broadening across our ensemble via a distribution of transition frequencies, which can be modelled as varying detuning, $\delta_i$, from a central transition with frequency $\omega_0=2 \pi \cdot f_0$. This detuning contributes to the total detuning $\Delta_i = 2 \pi \cdot \delta_i - (\omega _{MW} - \omega _0)$ between the transition frequency and the driving frequency $\omega_{MW}=2 \pi \cdot f_{MW}$. We define the term drive detuning to correspond to the difference $f _{MW} - f _0$ between the driving frequency $f_{MW}$ and the central transition frequency $f_0$. The drive detuning can be altered without changing $f_0$ by varying $f_{MW}$. We simulate microwave drive amplitude variations as an effective Rabi frequency $\alpha_i \Omega_R$ varying across the ensemble. Here, $\alpha_i$ is defined as the ratio between the effective Rabi frequency and the intended Rabi frequency in the non-detuned case $\Omega_R$, which is kept fixed~\cite{Poulsen21}. 

The ensemble is modelled as consisting of NV centers, each with a particular value of $\delta_i$ and $\alpha_i$, while all other parameters are assumed to be constant across the entire ensemble. A pump laser pulse is assumed to return every NV center to the $m_S=0$ ground state $\left|0 \right\rangle$ and the readout is assumed to reflect the exact state distribution. For simplicity, we consider only NV centers aligned along one crystallographic axis of diamond and assume that the magnetic field is aligned along that axis.
%The simulations are performed for each particular combination of $\delta_i$ and $\alpha_i$, and the results are averaged according to the assumed distributions of $\delta_i$ and $\alpha_i$. 
The $\delta_i$-values are considered to follow a Gaussian distribution centered at zero with a full width at half maximum ($\Gamma$). The values of $\delta_i$ are taken in the range $\pm \Gamma$. In our results and discussion below, the upper limit of this range is used to indicate the level of inhomogeneous broadening $L_{IHB}$ for a given simulation, i.e. a $\Gamma$ of 0.5 MHz equals $L_{IHB}=0.5$ MHz. The $\alpha_i$-values follow a flat distribution between 0 and 1 to approximate a decaying MW drive field. The range below 1 spanned by the $\alpha_i$-values is used to indicate the level of MW drive amplitude variations, termed $L_{DAV}$ below, i.e. if $\alpha_i$ varies between 0.9 and 1, $L_{DAV}=0.1$. MW drive amplitude variations can be neglected by setting $\alpha_i=1$ in all cases.

\subsection{Ramsey interferometry}
Simulations of Ramsey interferometry were performed using a Hamiltonian representing a coherently driven single NV two-level spin system in a rotating frame
\begin{equation} \label{eq:H}
    \hat{H} = \frac{\Delta_i}{2} \hat{\sigma} _z + \frac{\alpha _i \Omega_R}{2} \hat{\sigma} _x,
\end{equation}
where $\hat{\sigma}_x$ and $\hat{\sigma}_z$ are Pauli spin-matrices, and $\hbar$ is equal to one. The simulations are performed in three steps. First, an input state representing an NV center initialized in state $m_s=0$, $\ket{\psi _i} = \ket{0} = (1 \, 0 )^T$, is allowed to evolve under the influence of $\Omega_R$ for a duration $T_p = \pi / (2 \Omega _R)$ corresponding to and simulating the effect of a $\pi/2$-pulse. In the second step, the resulting state is then used as the input state for an evolution where $\Omega_R = 0$ for a duration equal to the considered free precession time $\tau$. In the final and third step, the resulting output state after free precession is used as the input state for an evolution where $\Omega_R$ is again non-zero for a duration of $T_p = \pi / (2 \Omega _R)$, accounting for the effect of the final $\pi/2$-pulse. For the output of our simulation, we define a normalized contrast $C_i$, obtained for the resulting final output state $\ket{\psi _f}$ by determining its overlap with state $\ket{1} = ( 0 \, 1 )^T$ corresponding to either $m_s = 1$ or $m_s = -1$,
\begin{equation} \label{eq:C}
    C_i = | \braket{1 |\psi _f}|^2.
\end{equation}

The obtained value of $C_i$ represents normalized fluorescence contrast between on and off microwave resonance, calculated for a single NV center in the ensemble with parameters $\Delta_i$ and $\alpha_i$.
The parameter $C_i$ is a valid measure of the normalized contrast, given that the contrast will be minimal when the NV center is in state $\ket{0}$ and maximal when the NV center is in state $\ket{1}$. We assume that a normalized contrast of $C_i=1$ corresponds to a fluorescence contrast of $30\%$, the maximum typically measured in an optically detected magnetic resonance experiment using a single NV center in bulk diamond~\cite{Jelezko2006}. We define a total normalized contrast for the ensemble $C$ as the mean normalized contrast taken by averaging across all $C_i$ values. 

We can also include pure dephasing in the simulation by using a quantum master equation for a two-level system
\begin{equation} \label{eq:master}
    \frac{d\rho}{dt} = -i \left[ \hat{H}, \rho \right] + \Gamma_{pure} \big( \hat{\sigma} _z \rho \hat{\sigma} _z - \frac{1}{8} \hat{\sigma} _z \hat{\sigma} _z \rho - \frac{1}{8} \rho \hat{\sigma} _z \hat{\sigma} _z \big),
\end{equation}
where $\rho = \ketbra{\psi}{\psi}$ for a pure state and the rate of pure dephasing is given by $\Gamma_{pure} \approx 1 / T_2^*$. Otherwise, the simulations are performed using the exact same three-step approach that was previously described for the case without pure dephasing. In density matrix notation, the input state corresponding to an NV initialized in state $m_s=0$ is $\rho_i = \begin{bmatrix} 1 & 0 \\ 0 & 0 \end{bmatrix}$, and the normalized contrast $C_i$ is directly obtained from the second diagonal element of $\rho _f$, which is the output density matrix obtained from the third and final step of the simulation. 

\begin{figure}
    \centering
    \includegraphics{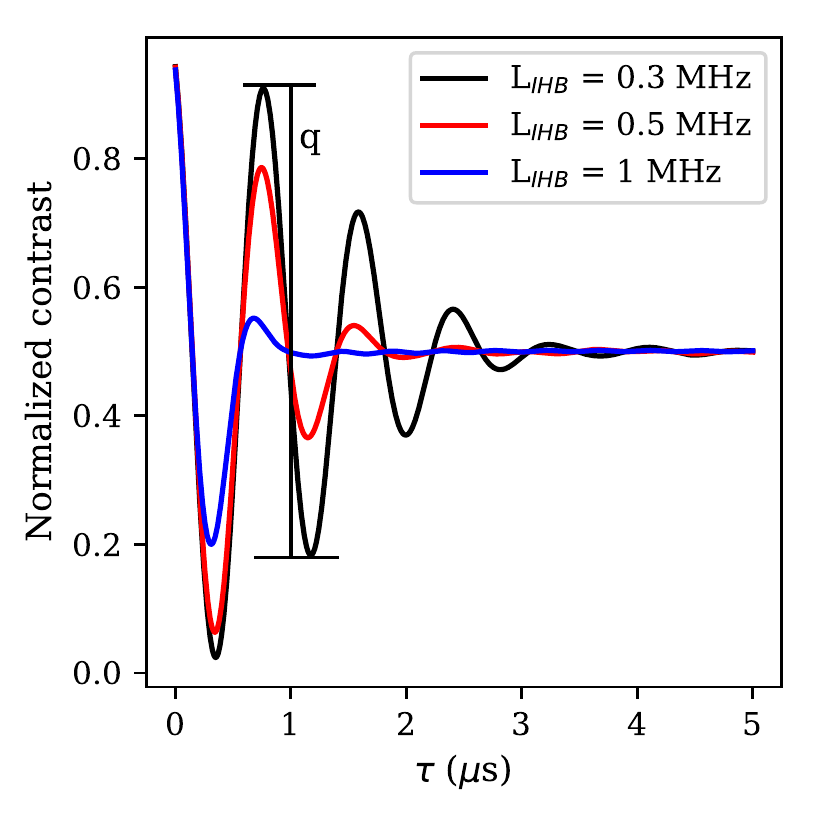}
    \caption{Color online. Simulated Ramsey measurement for $\Omega_R =2 \pi \cdot 5$ MHz, no pure dephasing, a drive detuning equal to 1.2 MHz and three different values of $L_{IHB}$. The quantity $q$ indicated on the plot for $L_{IHB}=0.3$ MHz is a self-defined figure of merit used to judge the relative strength of the Ramsey fringes.}
    \label{fig:Ramsey_ex1}
\end{figure}

As a demonstration of the effect of inhomogeneous broadening, we present in Fig.~\ref{fig:Ramsey_ex1} the ensemble-average normalized contrast $C$ as a function of free precession time simulated for the Ramsey protocol for a $N=41$ ensemble with $\Omega_R =2 \pi \cdot 5$ MHz, no pure dephasing, a drive detuning equal to 1.2 MHz, no drive amplitude variations and three different values of $L_{IHB}$.
As expected, we observe Ramsey fringes where the contrast is seen to oscillate at a frequency equal to the detuning between the drive and the central transition frequency. The Ramsey interference fringes seen in Fig.~\ref{fig:Ramsey_ex1}, are observed to decay within a time constant that roughly corresponds to $1/L_{IHB}$. The observed decay caused by inhomogeneous broadening is thus similar in effect to a decay of the contrast value caused by pure dephasing.

In order to provide a quantitative measure of the presence of Ramsey fringes and their relative strength, we define a figure of merit $q$ that is equal to the difference in normalized contrast between the second-highest fringe maximum and the second-lowest minimum, as indicated in Fig.~\ref{fig:Ramsey_ex1}. This ensures the presence of at least two clear fringes for large $q$, separating a genuine Ramsey interference measurement from relaxation or ensemble $\pi$-pulse rotation effects that might resemble a single fringe. For each value of $\Omega_R$ that we consider in the simulations, we then determine $q$ for different values of the drive detuning and $L_{IHB}$. For each set of $L_{IHB}$, we then extract the drive detuning that yields the maximum value $q_{max}$ and hence the most pronounced Ramsey fringes. The extracted drive detuning value is used for the estimation of the maximum achievable slope in normalized contrast for the set values of $\Omega_R$ and $L_{IHB}$.
\begin{figure}
    \centering
    \includegraphics{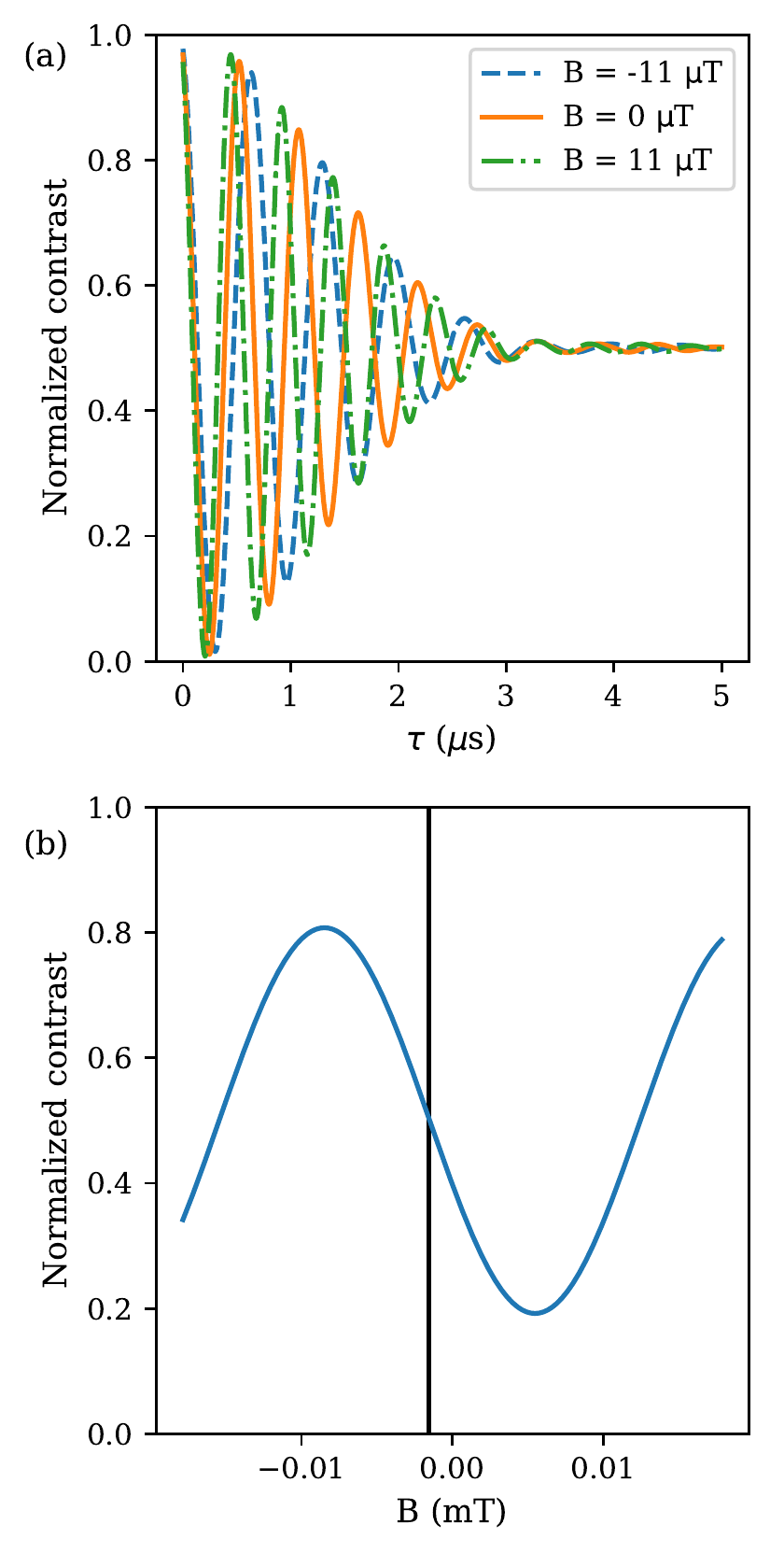}
    \caption{(a) Ramsey interferometry contrast versus $\tau$ for $L_{IHB}=0.3$ MHz, $\Omega_R=2\pi \cdot 10$ MHz, no pure dephasing and a drive detuning of 1.8 MHz with three different values of $B$. (b) Ramsey interferometry contrast versus $B$ for $\tau = 1.25 \, \upmu \si{s}$ and otherwise identical parameters to (a). Position of maximum slope $dC/dB$ indicated by vertical line.}
    \label{fig:Ramsey_principle}
\end{figure} 

For direct comparison with CW and $\pi$-pulse ODMR, it was necessary to convert the change in normalized contrast as a function of the target sensing parameter into the equivalent normalized contrast change (slope) as a function of variation in microwave drive frequency $f_{MW}$, $C^{'}$=$dC$/$df_{MW}$.
Unlike CW or $\pi$-pulse ODMR, where sensing is simply dependent on the change in microwave resonance frequency of the ODMR features caused by a change in the parameter to be sensed, e.g. magnetic field or temperature, the sensitivity of the Ramsey scheme approach is not directly related to the maximum slope of the Ramsey fringes. Instead, a change in the parameter to be sensed will change the resonance frequency, thereby changing the total detuning $\Delta_i$ and thus the oscillation frequency of the Ramsey fringes. This causes the Ramsey fringes to be compressed or stretched in free precession time $\tau$ and leads to a measurable oscillation in the contrast for a fixed $\tau \neq 0 $ as a function of the sensing parameter. This is exemplified in Fig.~\ref{fig:Ramsey_principle} for the case of magnetic field sensing. Here, the addition of a magnetic field $B$ shifts the resonance frequency and thus the total detuning by $m_s \gamma_e B$, with $\gamma_e = \SI{28}{MHz/mT}$ the gyromagnetic ratio of an electron spin. Figure \ref{fig:Ramsey_principle}(a) shows how the Ramsey fringes are shifted by a change in the magnetic field $B$ while Fig.~\ref{fig:Ramsey_principle}(b) illustrates how the contrast at a fixed $\tau$ oscillates as a function of the magnetic field $B$. The maximum slope of this oscillation $dC/dB$, shown in Fig.~\ref{fig:Ramsey_principle}(b) for $\tau=1.25 \, \upmu \si{s}$, is the point of maximum attainable sensitivity $\eta_B$ of Ramsey sensing of magnetic fields. The maximum achievable slope is dependent on the fixed value of $\tau$ due to the effects of inhomogeneous broadening and/or pure dephasing.

In order to determine the maximum sensitivity of the Ramsey scheme, it was therefore necessary to simulate the normalized contrast as a function of the sensing parameter (e.g. magnetic field $B$) for different fixed $\tau$, obtaining a $\tau_{opt}$ that maximises $\eta_B$. The maximum achievable slope as a function of the sensing parameter can then be converted to $C'$=$dC$/$df_{MW}$ using a conversion factor for the known change in transition frequency as a function of a given sensing parameter for a single NV axis (e.g. $\gamma_e=28$ MHz/mT for magnetic field or $-74.2$ kHz/K for temperature).

In this work, we performed this conversion by determining the change in ensemble average normalized contrast in a simulated applied magnetic field between $-3.6 \, \upmu \si{T}$ and $3.6 \, \upmu \si{T}$ with varying $\tau$ up to $5 \, \upmu \si{s}$. In each case considered in the simulation, the driving frequency $\omega_{MW}$ was chosen such that the total drive detuning $\Delta_i$ without field ($B=0$) maximizes the figure of merit $q$ to ensure clear Ramsey fringes. From these simulations we determined the maximum $dC/dB$ and converted this to $C'$=$dC$/$df_{MW}$ using the above conversion factor. This permitted direct comparison of the Ramsey interferometery scheme with CW and $\pi$-pulse ODMR sensing.

\subsection{$\pi$-pulse ODMR}
For the case of simulating the $\pi$-pulse ODMR scheme, simulations were performed using the same methods as for simulating the Ramsey protocol, but by replacing the three-step sequence with a single step where an input state $\ket{\psi _i} = \ket{0}$ was allowed to evolve under the influence of a drive field with strength $\Omega_R$ for a duration $T_p = \pi / \Omega _R$ equal to a $\pi$-pulse. The normalized contrast $C$ obtained for the resulting output state $\ket{\psi _f}$ was then calculated using Eq.~\ref{eq:C}. This approach was repeated for varying values of the drive frequency $f_{MW}$, from which a normalized contrast spectrum and the measurable maximum slope was extracted. \\

\subsection{CW ODMR}
We performed the simulation of the CW ODMR protocol using a five-level model of the NV center~\cite{Ahmadi2017}, including spin, optical and non-radiative transitions. The decay rates typical for an NV center in bulk diamond are listed in~\cite{Ahmadi2017SI}, and the optical excitation rate is $\Gamma_p$.

The steady-state solutions for the ground state populations for a single NV can be obtained directly via the expressions~\cite{Ahmadi2017}
\begin{equation} \label{eq:first}
    \rho_{11}^{ss} = \left[ 1 + \Xi + \frac{\Gamma_p}{K_3} + \frac{\Gamma_p \Xi}{K_4} + \frac{k_{35} \Gamma_p}{K_3 K_5} + \frac{k_{45} \Gamma_p \Xi}{K_4 K_5} \right]^{-1},
\end{equation}
and
\begin{equation}
    \rho_{22}^{ss} = \left[ 1 + \frac{1}{\Xi} + \frac{\Gamma_p}{K_4} + \frac{\Gamma_p}{K_3 \Xi} + \frac{k_{45} \Gamma_p}{K_4 K_5} + \frac{k_{35} \Gamma_p}{K_3 K_5 \Xi} \right]^{-1},
\end{equation}
where
\begin{equation}
    \Xi = \frac{\left[ \frac{k_{21}}{2} + \frac{\Gamma_p(k_{32}K_5+k_{52}k_{35})}{K_3 K_5} + \frac{(\alpha_i \Omega_R)^2 \gamma_2'}{2(\gamma_2'^2 + \Delta_i^2)} \right]}{\left[ \Gamma_p + \frac{k_{21}}{2} - \frac{\Gamma_p(k_{42}K_5 + k_{52}k_{45})}{K_4K_5} + \frac{(\alpha_i \Omega_R)^2 \gamma_2'}{2(\gamma_2'^2 + \Delta_i^2)} \right]}
\end{equation}
and
\begin{equation}
\begin{split}
    K_3 &= k_{35} + k_{31} + k_{32}, \\ K_4 &= k_{41} + k_{42} + k_{45}, \\ K_5 &= k_{51} + k_{52},
\end{split}
\end{equation}
where $\Gamma_p$ is the optical pumping rate, $k_{nm}$ is the decay rate from level $n$ to level $m$, $K_n$ is the total decay rate from level $n$, $\gamma_2' = \gamma_2 + \Gamma_p / 2$ is the optical dephasing rate, and $\gamma_2 = 2 \pi / T_2^* + k_{21} / 2$ is the spin dephasing rate.
The steady-state solutions for the ground state populations can be used to obtain the CW ODMR spectrum as
\begin{equation}
    \mathcal{I}_{CW} = \beta_3 \rho_{33}^{ss} + \beta_4 \rho_{44}^{ss} = \beta_3 \frac{\Gamma_p}{K_3} \rho_{11}^{ss} + \beta_4 \frac{\Gamma_p}{K_4} \rho_{22}^{ss}
\end{equation}
where
\begin{equation} \label{eq:last}
    \beta_3 = \frac{k_{31} + k_{32}}{K_3}, \: \beta_4 = \frac{k_{41} + k_{42}}{K_4}.
\end{equation}
We include pure dephasing via $2 \pi /T_2^*$ in the term $\gamma_2' = 2 \pi / T_2^* + k_{21} / 2 + \Gamma_p / 2$ and can neglect the influence of pure dephasing by setting $2 \pi/T_2^* = 0$. Eqs. (\ref{eq:first}-\ref{eq:last}) can thus be used to obtain the CW ODMR spectrum for a single NV center as a function of microwave drive frequency.

The spectrum values $\mathcal{I}_{CW}$ were converted to contrast relative to the value obtained when the microwave drive is far off-resonance. These values are divided by the maximum contrast obtained in the asymptotic limit where $\Omega_R$ is very large in order to obtain a normalized contrast value $C_i$. By repeated simulation for all values of $\Delta_i$ and $\alpha_i$, we then calculate the ensemble average normalized contrast $C$ as the mean of all $C_i$ values in the ensemble. By varying the simulated microwave drive frequency $f_{MW}$, we can then derive the maximum slope as a function of drive frequency $C^{'}$=$dC$/$df_{MW}$. This can be directly compared to the equivalent values obtained from the simulation of a Ramsey or $\pi$-pulse scheme, representing a measure of the maximum achievable sensitivity for CW ODMR sensing.

\section{Results and discussion}
\subsection{Ramsey interferometry}
\begin{figure}
    \centering
    \includegraphics{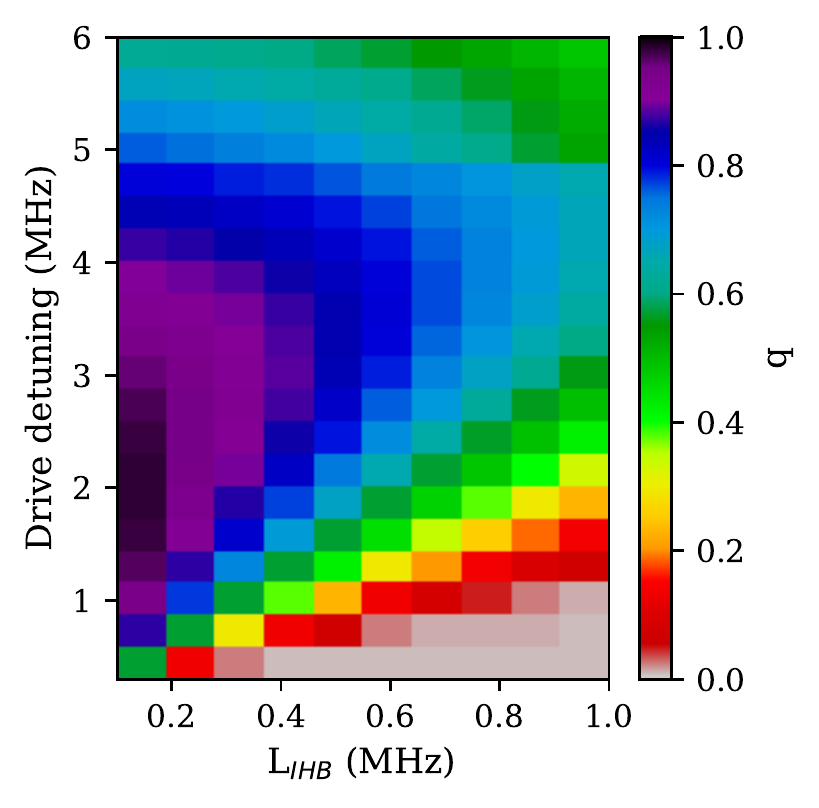}
    \caption{Plot of the self-defined figure of merit $q$ as a function of drive detuning and $L_{IHB}$ for $\Omega_R =2 \pi \cdot 3$ MHz and no pure dephasing or drive amplitude variations.}
    \label{fig:Ramsey_qR3}
\end{figure}

We first calculated how the visibility of the Ramsey fringes depended on the degree of drive detuning and the inhomogenous broadening ($L_{IHB}$). This was done in terms of the figure of merit $q$, to give a clear picture of how well the interferometry performs as a function of these parameters. Figure \ref{fig:Ramsey_qR3} shows a plot of $q$ as a function of drive detuning and inhomogenous broadening at a fixed overall Rabi frequency $\Omega_R =2 \pi \cdot 3$ MHz with no pure dephasing or drive amplitude variations included. In the Supplementary information, we include examples of the individual Ramsey interferometry simulations that were used to obtain this plot. It is clear that the optimal drive detuning to maximize the figure of merit $q$ and thus the quality of the Ramsey fringes increases with increasing inhomogeneous broadening. Moreover, we see that the fringes worsen with increasing inhomogeneous broadening. This is as expected, with an increase in $L_{IHB}$ leading to a faster decay of the Ramsey fringes and thus requiring a larger oscillation frequency (drive detuning) in order to obtain sufficient Ramsey fringes to perform Ramsey interferometry. Also as expected, as $L_{IHB}$ reaches its upper range ($>1$ MHz), it becomes increasingly difficult to obtain interference fringes, as the broadening means the fixed microwave pulse length is no longer $\pi/2$ for an increasingly larger number of NV center spins in the ensemble.

By the procedure outlined in Methods, we calculate the maximum fluorescence contrast slope $C'$ for Ramsey interferometry. We initially focus on a particular scenario for an ideal high sensitivity measurement with a strong, uniform microwave field across the diamond resulting in no ensemble drive amplitude variations and a high Rabi frequency ($\Omega_R > 2 \pi \cdot 2$ MHz). We then simulate the influence of Rabi frequency and inhomogeneous broadening on the predicted maximum slope $C'$. 

\begin{figure}
    \centering
    \includegraphics{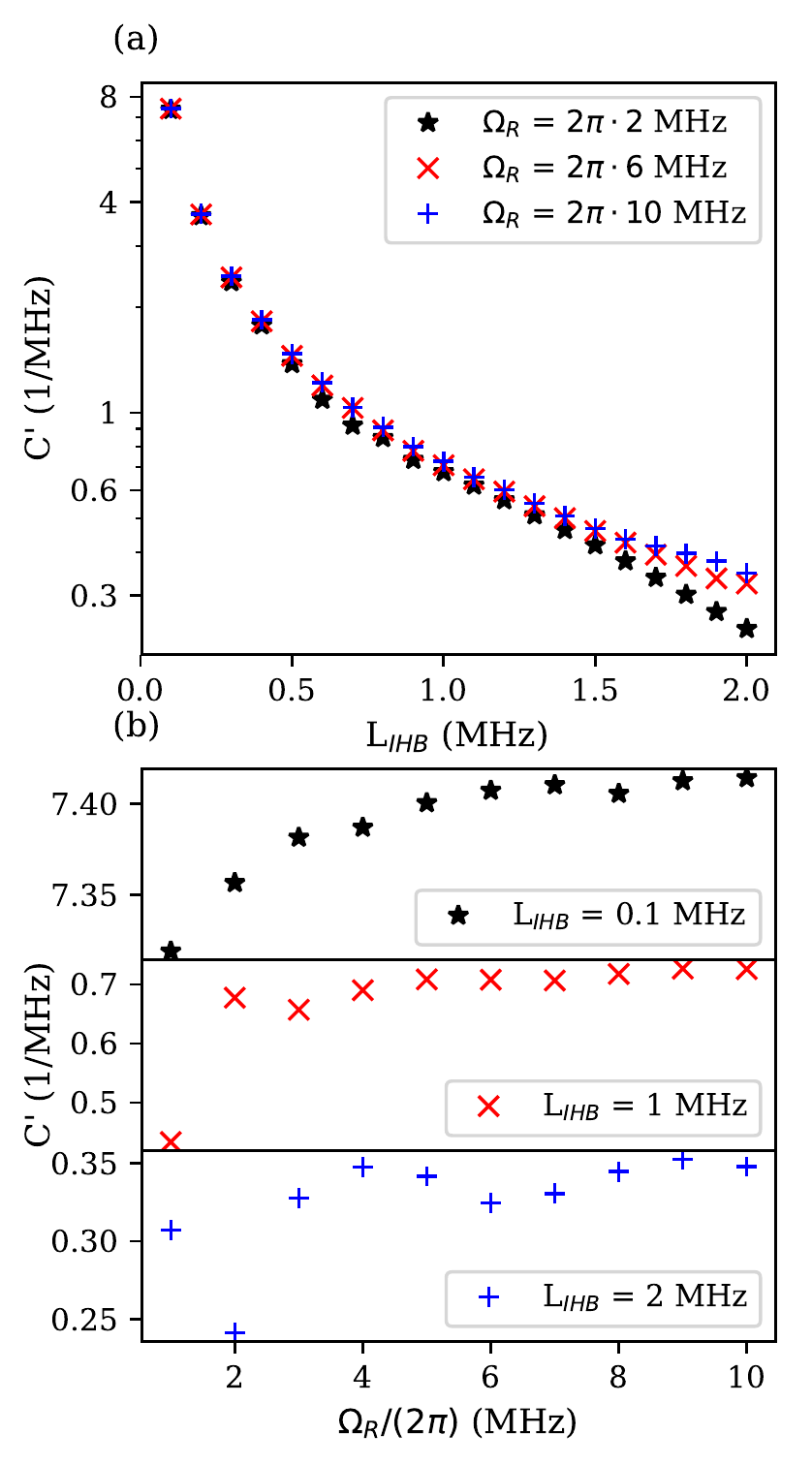}
    \caption{Maximum achievable slope in normalized contrast for Ramsey interferometry with no drive amplitude variations or pure dephasing as a function of (a) inhomogeneous broadening with various Rabi frequencies and (b) Rabi frequency with various levels of inhomogeneous broadening.}
    \label{fig:Ramsey_S_vs_IHB}
\end{figure} 

Figure \ref{fig:Ramsey_S_vs_IHB}(a) shows the simulated $C'$ as a function of inhomogeneous broadening for three different Rabi frequencies and Fig.~\ref{fig:Ramsey_S_vs_IHB}(b) shows the simulated $C'$ as a function of Rabi frequency for three different $L_{IHB}$-values, with $L_{DAV}$=0 (no drive amplitude variations) and no pure dephasing. The plots in Fig.~\ref{fig:Ramsey_S_vs_IHB}(b) also include anomalous features where the slope decreases with increasing Rabi frequency. This arises because for each value of $\Omega_R$ and $L_{IHB}$, we calculate the drive detuning that yields the maximum $q$, rather than maximizing $C'$ directly. As outlined in \textit{Methods}, this is to maximize fringe visibility to remain in the regime where we are performing genuine Ramsey interferometry. If this were not the case, the optimization routine might, for example, find the maximum $C'$ for zero or near-zero precession time $\tau$, which would be indistinguishable from the $\pi$-pulse scheme. The result of the method we choose is that the optimized drive detuning between values of $\Omega_R$ or $L_{IHB}$ may reach different local maxima in $q$, leading to a step change in $C'$ between adjacent points. This can lead to the anomalies we observe. 

Two aspects are clear from the data shown in Fig.~\ref{fig:Ramsey_S_vs_IHB}(a). First, the achievable slope $C'$ in the Ramsey scheme decreases in a pseudo-exponential fashion with increasing inhomogeneous broadening. This indicates a lower sensitivity for all schemes with higher inhomogeneous broadening. As discussed above, this is what would be expected based on the observed effect of inhomogeneous broadening, acting to reduce the fringe visibility and number as illustrated in Fig.~\ref{fig:Ramsey_ex1}. Secondly, the achievable slope increases only slightly with increasing Rabi frequency, particularly at low inhomogeneous broadening. This result is somewhat surprising, as it would be expected that by increasing the microwave power supplied using a pulsed scheme resistant to power broadening, the microwave resonance should be better defined, increasing contrast $C$ and $C'$. 

The slight increase can be attributed to the relationship between Rabi frequency and $\pi/2$-pulse performance. As Rabi frequency is increased, the $\pi/2$-pulses will be able to perform well for NV center spins with larger detuning from the driving frequency. However, once the pulses are able to perform well for most or all of the target ensemble, increasing the Rabi frequency further does not significantly improve the overall pulse performance and thus does not significantly improve the achievable slope. The point where the $\pi/2$-pulses are able to affect most of the ensemble correctly naturally occurs at a lower Rabi frequency for ensembles with lower levels of inhomogeneous broadening. 

This result implies that for sensing experiments using the Ramsey scheme, increasing microwave amplification or enhancing antenna design to achieve a higher Rabi frequency will not necessarily lead to significant enhancement in sensitivity. This is an important result for both benchtop systems aiming for state of the art sensitivity and for systems with low Rabi frequency, such as microfabricated NV sensors with low available microwave power~\cite{Zhu2017, Mi2020}. For a high quality diamond with low $L_{IHB}=0.1$ MHz, an increase in Rabi frequency from 1 MHz to 10 MHz only increases the slope $C'$ by a factor of $\approx$ 1.3$\%$, see Fig.~\ref{fig:Ramsey_S_vs_IHB}. In our zero noise model, this corresponds to the same factor of enhancement in sensitivity to any target parameter using this scheme. 

\begin{figure}
    \centering
    \includegraphics{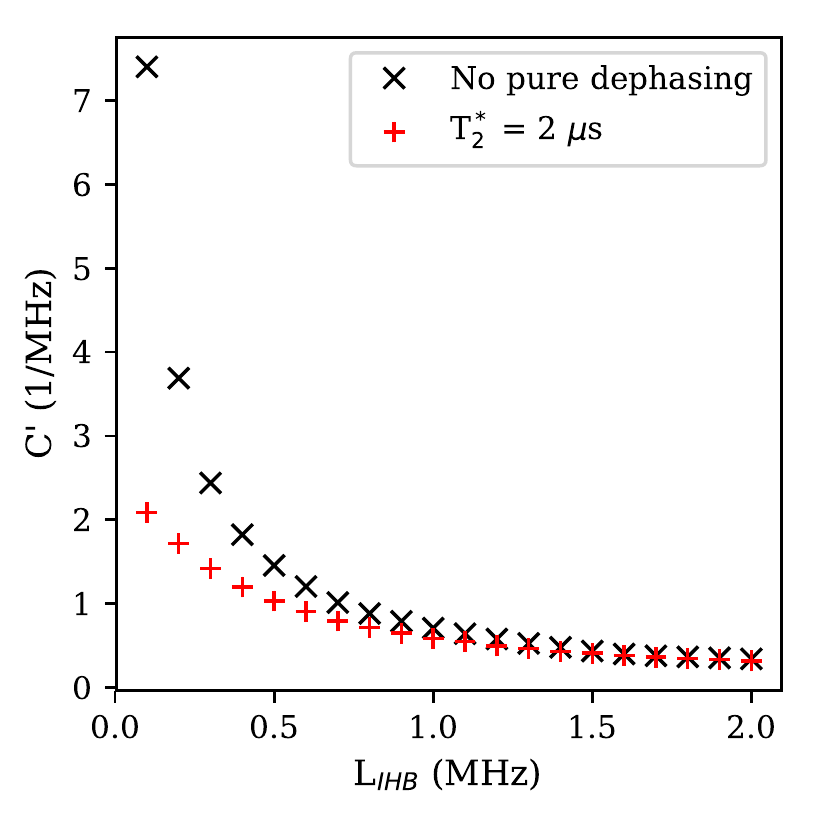}
    \caption{Maximum achievable slope in normalized contrast as a function of inhomogeneous broadening for Ramsey measurements with $\Omega_R =2 \pi \cdot 5$ MHz with and without pure dephasing. The indicated $T_2^*$-value is the one induced by pure dephasing alone.}
    \label{fig:Ramsey_comp1}
\end{figure}

We also simulate the introduction of pure dephasing, which additionally acts to reduce the Ramsey fringe visibility. This has an effect very similar to that of inhomogeneous broadening (Fig.~\ref{fig:Ramsey_principle}), with a decay in normalized contrast. The limit on interferometry performance will therefore depend on both pure dephasing and inhomogeneous broadening and their relative strengths defined by $T_2^{*}$ and $L_{IHB}$, respectively. We illustrate this in Fig.~\ref{fig:Ramsey_comp1} where the simulated $C'$ as a function of inhomogeneous broadening is plotted with and without pure dephasing.

The difference between the slopes obtained decreases with higher inhomogeneous broadening. In the regime of large values of $L_{IHB}$, the maximum achievable slope is limited by the inhomogeneous broadening. For low values of $L_{IHB}$, the maximum $C'$ is constrained by pure dephasing. This implies that even if long $T_2^{*}$ times are achieved, as is often sought in sensing experiments through well designed pulsed protocols, the inhomogeneous broadening will still act to ultimately constrain the maximum sensitivity of the Ramsey scheme, and that increasing the Rabi frequency (through increasing the MW power) of the MW pulses will not necessarily overcome this issue. This highlights the critical importance of material (diamond) design to minimize IHB (e.g. by minimizing inhomogeneous non-NV diamond nitrogen content or minimizing material strain). 

\subsection{CW and $\pi$-pulse ODMR}
\begin{figure}
    \centering
    \includegraphics{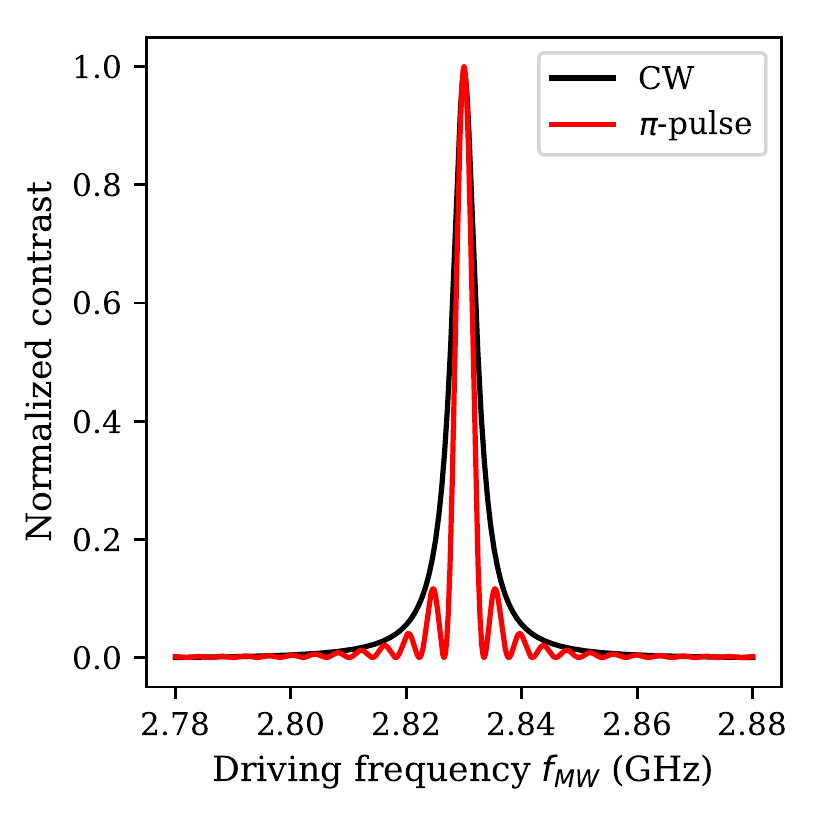}
    \caption{Normalized contrast values for a simulated CW ODMR spectrum for $\Gamma_p = 2 \pi \cdot 1$ MHz, $\Omega_R =2 \pi \cdot 2$ MHz and a simulated $\pi$-pulse ODMR spectrum for $\Omega_R =2 \pi \cdot 2$ MHz with zero inhomogeneous broadening, no pure dephasing and no MW drive amplitude variations using the five-level model.}
    \label{fig:CWcontrast1}
\end{figure}

Fig.~\ref{fig:CWcontrast1} shows example simulated ODMR spectra in terms of normalized ensemble contrast $C$ as a function of microwave drive frequency $\omega_{MW}$ for CW and $\pi$-pulsed schemes. We show a single resonance feature corresponding to a single NV axis, split by 40 MHz by a simulated static magnetic field with a 1.4 mT amplitude along the NV axis. The parameters we use are $\Gamma_p = 2 \pi \cdot 1$ MHz, $\Omega_R =2 \pi \cdot 2$ MHz for the CW simulation and $\Omega_R =2 \pi \cdot 2$ MHz for the $\pi$-pulse simulation. We plot the ideal spectra, with zero pure dephasing, inhomogeneous broadening and no MW drive amplitude variations. The $\pi$-pulse ODMR spectrum was obtained using a fixed $\pi$-pulse duration chosen to match the on-resonance frequency. This leads to oscillations in contrast as this fixed duration periodically matches different Rabi frequencies obtained at different driving frequencies. Examples of simulated spectra varying the above parameters are shown in the Supplementary Information. 

We note that for CW and $\pi$-pulse ODMR, the relationship between slope $C'$ and Rabi frequency is not straightforward and larger Rabi frequency can result in lower slope. This is due to the fact that while larger Rabi frequency will result in larger achievable ODMR contrast, which serves to increase the slope, the additional microwave power necessary to achieve it will broaden the NV resonance linewidths, which acts to decrease the slope $C'$. The inverse is also true for green laser intensity/pumping rate, where resonance linewidth can narrow with higher laser intensity \cite{Rui2018, Barry2020}. For the ODMR simulations, we thus individually optimize the Rabi frequency and the green pumping rate to yield the maximum slope for each $L_{IHB}$ and $L_{DAV}$.

\subsection{Scheme comparison}
We compare the maximum sensitivity for the Ramsey interferometry, CW and $\pi$-pulse ODMR sensing schemes. Again, we assume zero noise such that the sensitivity is directly proportional to the ensemble response via slope $C'$. For simplicity of comparison, we compare the best possible Ramsey interferometry simulations using $\Omega_R = 2 \pi \cdot 10$ MHz with the other two schemes. We consider that $\Omega_R = 2 \pi \cdot 10$ MHz is a reasonable upper limit that can be achieved experimentally~\cite{Nobauer2015, Rembold2020}.

\begin{figure}
    \centering
    \includegraphics{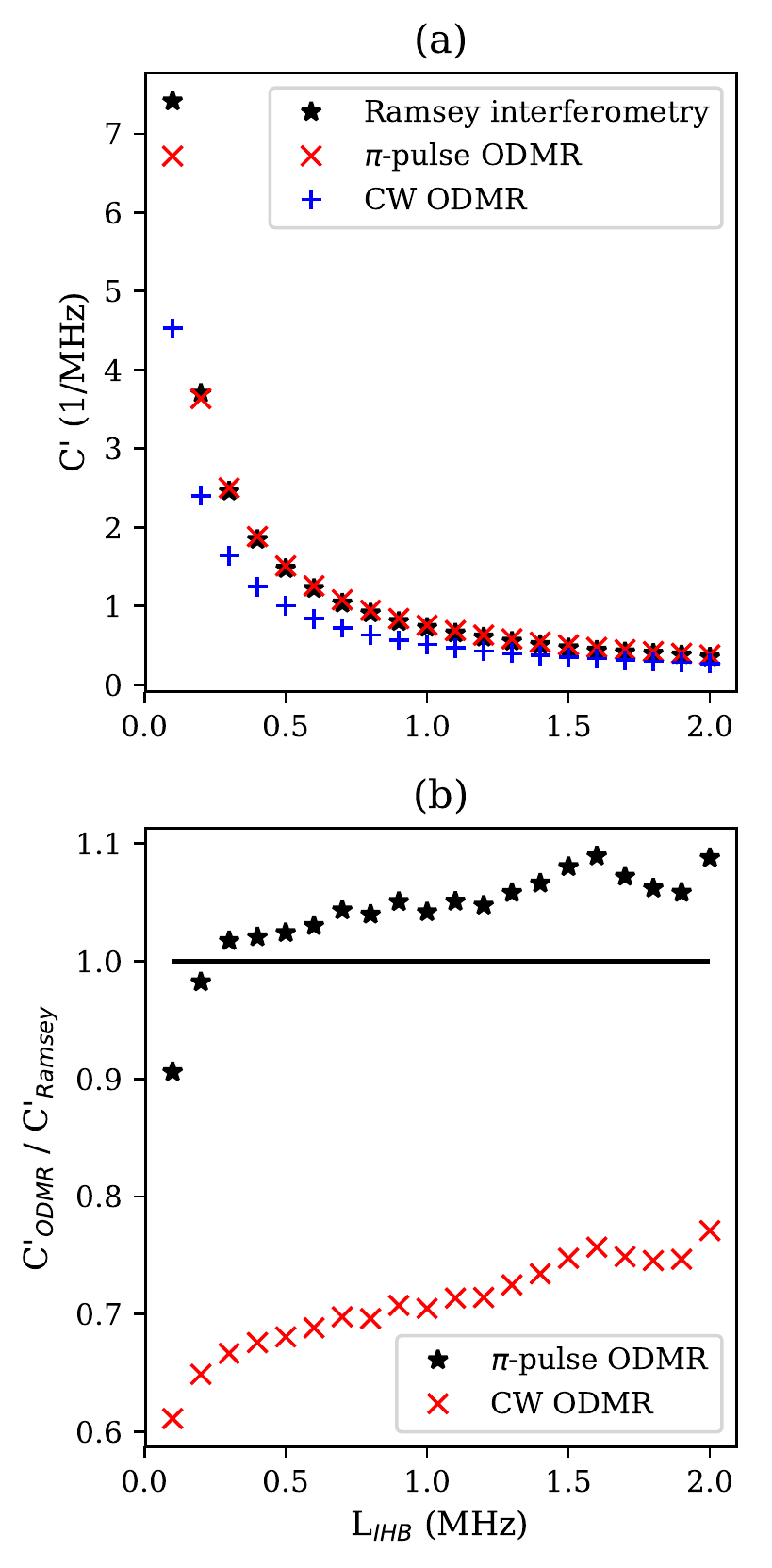}
    \caption{a) Maximum achievable slope in normalized contrast as a function of inhomogeneous broadening for Ramsey measurements with $\Omega_R =2 \pi \cdot 10$ MHz, optimized CW ODMR and optimized $\pi$-pulse ODMR. All values were obtained while neglecting drive amplitude variations and pure dephasing. b) Ratios between the ODMR and Ramsey interferometry slopes shown in a).}
    \label{fig:Ramsey_vs_ODMR_noa}
\end{figure}

We first consider only the effects of inhomogenous broadening, while neglecting drive amplitude variations and pure dephasing. Figure \ref{fig:Ramsey_vs_ODMR_noa}(a) shows the resulting simulated maximum slope as a function of $L_{IHB}$ for Ramsey interferometry, CW ODMR and $\pi$-pulse ODMR while neglecting drive amplitude variations and pure dephasing. The obtained optimal Rabi frequencies for CW and $\pi$-pulse ODMR were below $2 \pi \cdot 3$ MHz in all cases. For ease of comparison, the ratios between the obtained ODMR slopes and the obtained Ramsey interferometry slopes are shown in Fig.~\ref{fig:Ramsey_vs_ODMR_noa}(b) with a horizontal line at unity, i.e. equal performance.

The plots in Fig.~\ref{fig:Ramsey_vs_ODMR_noa} illustrate how the achievable slope decreases with increasing inhomogeneous broadening for all three techniques, but the decrease is most significant in the low IHB regime ($<0.2$ MHz) for Ramsey interferometry. As can be seen in Fig.~\ref{fig:Ramsey_vs_ODMR_noa}(b), the simulations predict a larger $C'$ and hence greater achievable sensitivity for $\pi-$pulse ODMR than Ramsey interferometry when $L_{IHB} \geq 0.3$ MHz. Given that $\pi$-pulse ODMR is no more complicated to implement than Ramsey interferometry, this indicates that $\pi$-pulse ODMR should be the preferred protocol when the inhomogeneous broadening is in this regime. For $L_{IHB} < 0.3$ MHz, the simulations predict the largest slope for Ramsey interferometry, indicating that one can gain sensitivity by implementing Ramsey interferometry in this regime. The CW ODMR simulations at best predict a slope that is close to $80\%$ of the Ramsey interferometry slope predicted for the same $L_{IHB}$. 

The plots in Fig.~\ref{fig:Ramsey_vs_ODMR_noa}(b) also include several anomalous features such as the ratio decreasing with increasing $L_{IHB}$ at several points and oscillations in the data for increasing $L_{IHB}$. As previously detailed, these arise due to the chosen method of optimizing the drive detuning. We can also perform this optimization directly in terms of $C'$ only, from which the same overall trends can be observed. This data can be seen in the Supplementary information. 

\begin{figure*}
    \centering
    \includegraphics{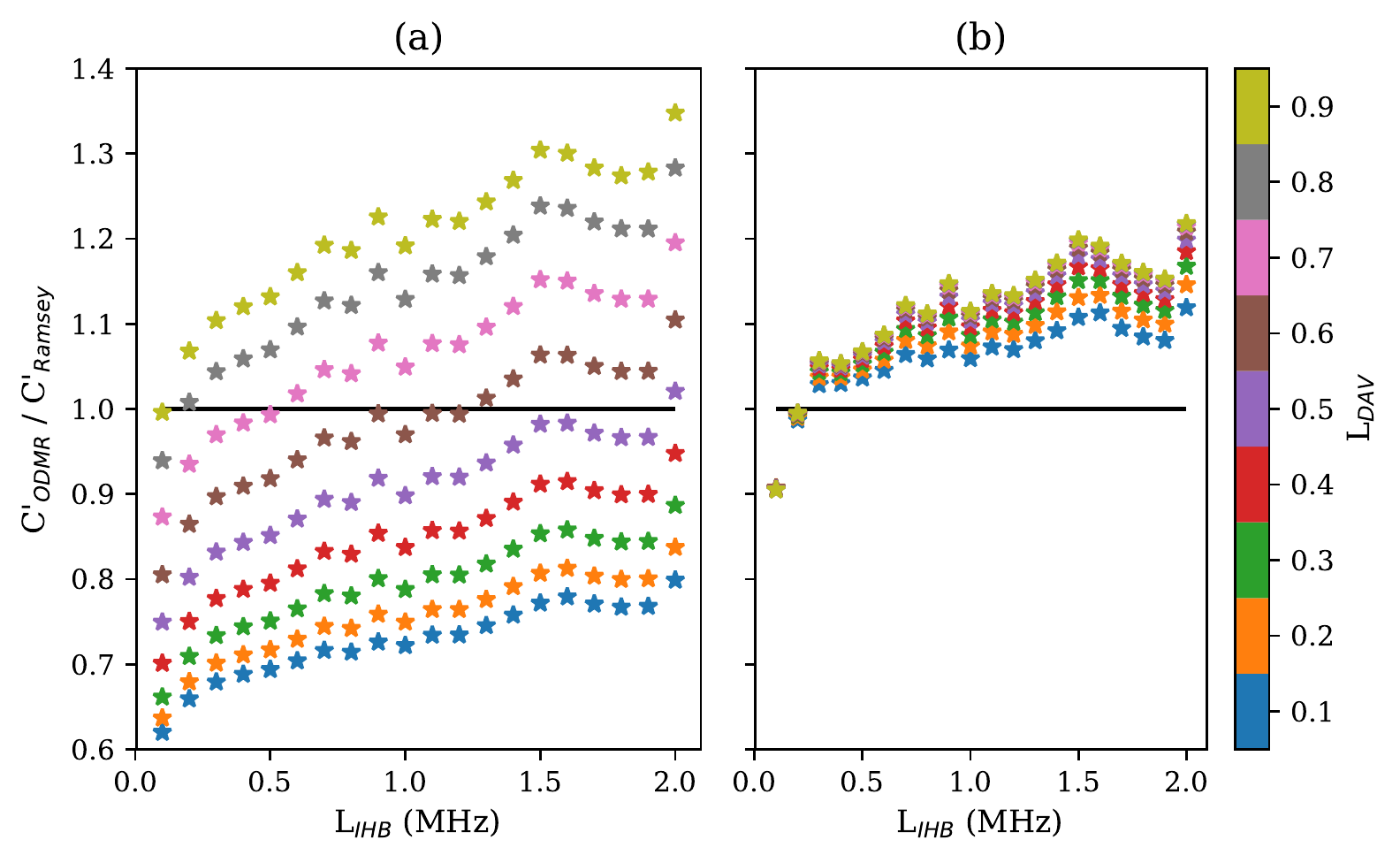}
    \caption{Ratio between the maximum achievable slope in normalized contrast for (a) optimized CW ODMR or (b) optimized $\pi$-pulse ODMR and Ramsey measurements with $\Omega_R =2 \pi \cdot 10$ MHz as a function of inhomogeneous broadening for varying levels of MW drive amplitude variations.}
    \label{fig:Ramsey_vs_ODMR_alpha}
\end{figure*}

Having investigated the behavior of the three sensing schemes in the absence of variations in microwave drive across the ensemble, we then perform the same comparison with finite drive amplitude variation $L_{DAV}$. Figure \ref{fig:Ramsey_vs_ODMR_alpha} shows plots of the ratio between the simulated maximum achievable slope for (a) CW ODMR and (b) $\pi$-pulse ODMR and Ramsey interferometry as a function of $L_{IHB}$ for MW drive amplitude variations between $10\%$ and $90\%$ ($L_{DAV}=0.1$-0.). Here we see that increasing the level of MW drive amplitude variation increases the performance (higher $C'$) of CW and $\pi$-pulse ODMR relative to Ramsey interferometry. For $\pi$-pulse ODMR, this effect is relatively small, with the same distinction in performance above and below $L_{IHB}=0.3$ MHz as simulated in Fig.~\ref{fig:Ramsey_vs_ODMR_noa}. However, the effect is far more pronounced for CW. When the MW drive amplitude variations exceed $50\%$, CW starts to outperform Ramsey measurements at large levels of inhomogeneous broadening. As the MW drive amplitude variations increase further, CW will outperform Ramsey at smaller and smaller levels of inhomogeneous broadening, eventually becoming preferable for nearly all levels of inhomogeneous broadening. For MW drive amplitude variations of around $80\%$ or larger, CW ODMR also outperforms $\pi$-pulse ODMR, as evidenced by achieving a larger ratio to the Ramsey measurement slopes. This is a surprising result, as pulsed readout schemes are usually assumed to offer better sensitivity (which is true in the limit of pure dephasing) \cite{Dreau2011}. Here we instead show there can be regimes where sensitivity is greater for a continuous wave scheme than for the pulsed alternatives when other factors are considered. 

We account for this result as arising due to the pulsed schemes being more adversely affected by MW drive amplitude variations. These schemes rely on the precise application of $\pi$ and $\pi/2$ pulses of the correct length for the maximum number of NV center spins in an ensemble. By increasing the variation in drive amplitude across the sample, this condition is no longer met for an increasingly large fraction of NV center spins. This leads to a reduction in both contrast and slope through broadening effects associated with incorrect length pulse application \cite{Wang2012}. For CW, this problem is less apparent as the states of the defect centers are continuously driven between the $m_S=0$ to $m_S=\pm 1$ levels, maintaining contrast as long as there is sufficient pump laser power to maintain spin polarization in the system. 

We consider this result to be particularly relevant for sensing schemes where such a wide variation in drive amplitude may occur. This includes two particular scenarios: 1) where the diamond is relatively large with respect to an antenna and the field of view of fluorescence collection is large and 2) where the microwaves are particularly localized to maximize contrast in a particular region. The first such scenario describes widefield imaging with NV centres, where there can be considerable variation in drive frequency across a diamond \cite{Mizuno2018, McCloskey2019}. The second may describe a confocal experiment, using lithographically patterned microwave antennas. 

Elements of both scenarios cover sensing experiments looking at nanodiamonds, for example for temperature sensing in biological tissue \cite{Simpson2017}. Here both $L_{DAV}$ and $L_{IHB}$ can be high due to variations in microwave drive if looking at multiple diamonds across a large single field of view and due to large differences in IHB between very few NV centers within each diamond, with considerable effects of crystal strain and surface interaction due to their small form factor \cite{Shames2015}.
\begin{figure}
    \centering
    \includegraphics{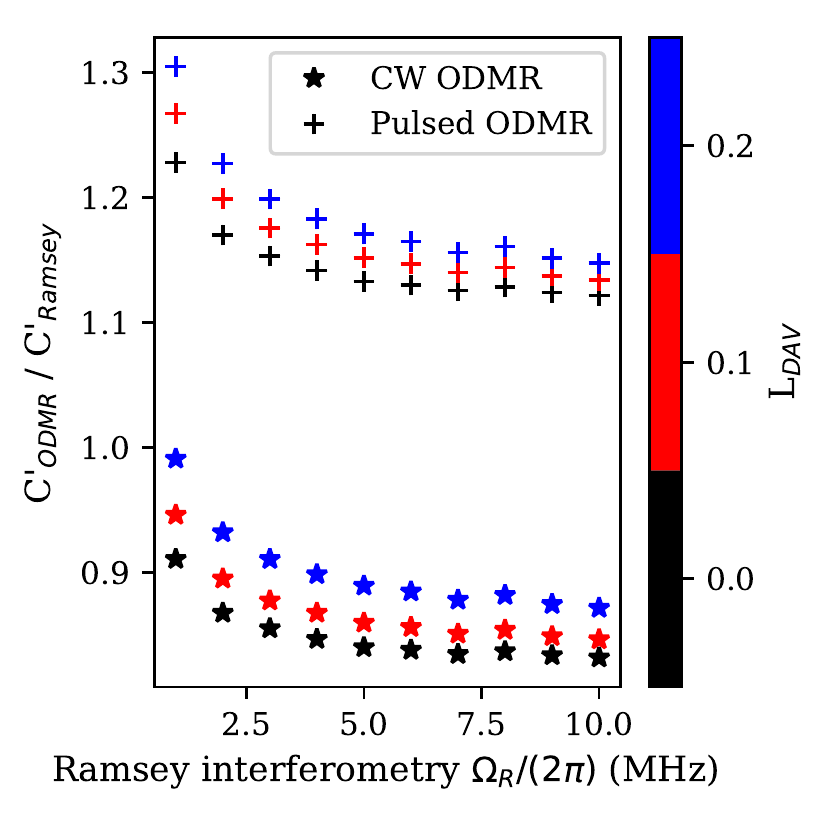}
    \caption{Ratio between the maximum achievable slope in normalized contrast for optimized CW or $\pi$-pulse ODMR and Ramsey measurements as a function of the Ramsey interferometry $\Omega_R$ for a measured $\delta_i$-distribution with varying levels of MW drive amplitude variations.}
    \label{fig:EnsdJim1inv}
\end{figure} 

Finally, we seek to perform our simulations using experimental data from a real diamond sample, to estimate the optimal sensing scheme for a real diamond. We achieve this through widefield CW ODMR imaging to extract the variation in zero offset magnetic field microwave resonance frequency across a 10$\times$10 $\upmu \si{m^2}$ region of a diamond. We use a chemical vapor deposition (CVD) diamond overgrown with $\ce{^15N}$ with $\ce{^12C}$ purification. The procedures and diamond details we use for the measurement are the same as given in \cite{Webb2021}. As this region is significantly smaller than our nearfield antenna, we assume that across the field of view we image there is a homogeneous microwave field, with minimal variation in microwave drive amplitude. We therefore assume that all of the variation we observe is due to inhomogenous broadening as our input $\delta_i$-distribution. This experimental distribution can be seen in Appendix \ref{App:dist}. Figure \ref{fig:EnsdJim1inv} shows the simulated ratios between the maximum achievable slope in normalized contrast for optimized CW or $\pi$-pulse ODMR and Ramsey interferometry as a function of the Rabi frequency $\Omega_{R}$ used for Ramsey interferometry with three different values of $L_{DAV}$. We calculate $\pi$-pulse ODMR to be the best possible scheme for all values of microwave drive variation, particularly in the low microwave power (low Rabi frequency) regime. Our simulations also suggest that in this regime, CW ODMR is close to outperforming the sensitivity of the Ramsey scheme. This diamond has recently been used for CW ODMR sensing of electrical current \cite{Webb2021}, but we do not presently have the capability to perform pulsed sensing using it. We also note that other factors such as readout noise and measurement time can play a role in a real experiment. Here we aim only to maximize the response of the sensing medium (the NV ensemble) in terms of maximizing the change in fluorescence output (slope $C'$) in response to a target factor (e.g. magnetic field or temperature). 

\section{Conclusion}
In this work, we compared Ramsey interferometry, $\pi$-pulse ODMR and CW ODMR sensing schemes using nitrogen-vacancy centers and investigated the impact of inhomogeneous broadening and MW drive amplitude variations. We demonstrate that the achievable response of the sensing ensemble in terms of the contrast slope $C'$, which is directly proportional to sensitivity, plateaus with increasing Rabi frequency for Ramsey interferometry. The performance of Ramsey interferometry can therefore not necessarily be improved simply by increasing the MW power. This is of considerable interest for realization of devices with low available power, such as microfabricated sensors with integrated semiconductor amplification, where significant MW amplification is not easy to realize due to e.g. heat dissipation concerns. 

We demonstrate that when the inhomogeneous broadening exceeds 0.3 MHz, $\pi$-pulse ODMR yields a larger $C'$ than Ramsey interferometry, indicating that $\pi$-pulse ODMR gives higher sensitivity in this regime. This indicates that Ramsey interferometry should only be considered when using a high-quality diamond with low inhomogeneous broadening, such as in low impurity samples with a relatively low nitrogen content~\cite{Wolf2015,Zhang2021}. 

The influence of MW drive amplitude variations is demonstrated to significantly improve the relative performance of CW ODMR compared to pulsed sensing schemes and $\pi$-pulse ODMR. For MW drive amplitude variations of around $80\%$ or larger, CW ODMR can outperform both Ramsey interferometry and pulsed ODMR for even a low level of inhomogeneous broadening. These results are of particular significance to applications that naturally have high levels of drive field variation, such as wide field of view NV center imaging \cite{Mizuno2018, McCloskey2019} or imaging of microwave frequency microcircuitry \cite{Webb2021}.

Finally, we apply our simulation to measurements from a real diamond, recently used for CW imaging \cite{Webb2021}. Our calculations predict that sensitivity could be improved by switching to a $\pi$-pulse sensing scheme. We however note that other factors beyond the scope of this work may limit sensitivity and these may vary between the schemes considered. In particular, in this work we only consider the maximal response of the sensing medium. We do not consider noise, arising from electronic or optical sources, or errors in pulse application. Further investigation beyond the scope of this work is required to fully examine these aspects before confirmation of our prediction.   

Our results represent an important step towards a greater understanding of the relation between sensor material properties, sensing regime and the ideal sensing scheme. We consider our results of particular interest to groups working with DC to low-frequency magnetometry and temperature sensing, particularly from biosamples, using diamonds both off-the-shelf or irradiated where low inhomogeneous broadening cannot be guaranteed.

\begin{appendices}
\section{Measured $\delta_i$-distribution} \label{App:dist}
Plot of the experimentally measured $\delta_i$-distribution used for the simulations in Fig.~\ref{fig:EnsdJim1inv}. The distribution was measured on a CVD diamond overgrown with $\ce{^15N}$ without isotopic purification.
\begin{figure}
    \centering
    \includegraphics{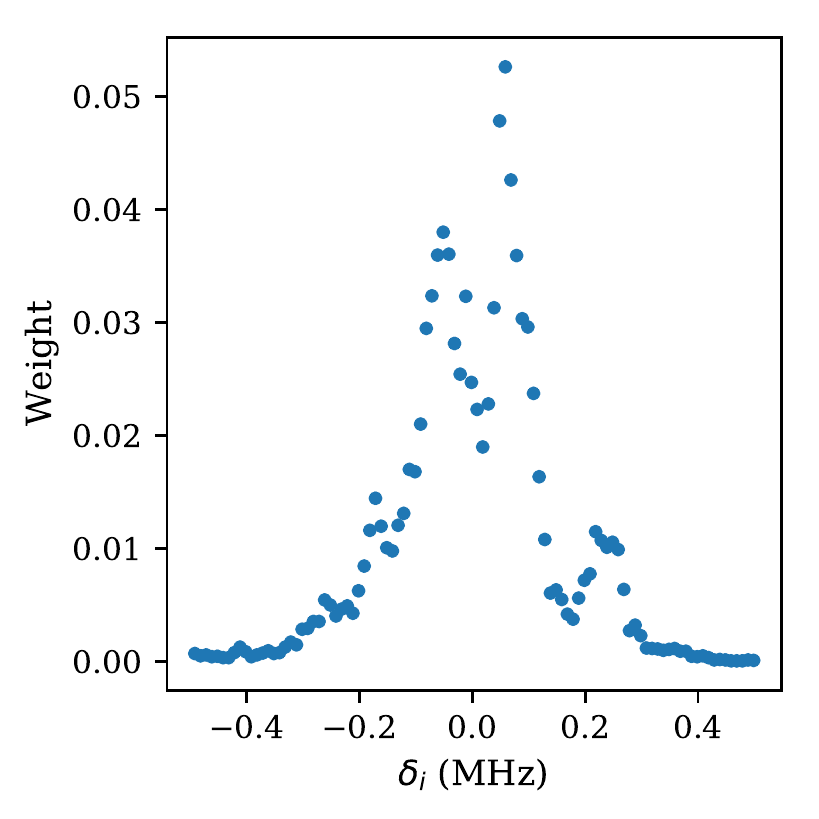}
    \caption{Experimentally measured $\delta_i$-distribution from a CVD diamond overgrown with $\ce{^15N}$ without isotopic purification. The distribution was used for the simulations in Fig.~\ref{fig:EnsdJim1inv}.}
    \label{fig:my_label}
\end{figure}

\end{appendices}

\bibliographystyle{unsrt}
%\bibliography{Literature}

\begin{thebibliography}{10}

\bibitem{Schroder2016}
T.~Schr\"{o}der, S.~L. Mouradian, J.~Zheng, M.~E. Trusheim, M.~Walsh, E.~H.
  Chen, L.~Li, I.~Bayn, and D.~Englund.
\newblock Quantum nanophotonics in diamond.
\newblock {\em J. Opt. Soc. Am. B}, 33(4):B65--B83, 2016.

\bibitem{Aharonovich2016}
I.~Aharonovich, D.~Englund, and M.~Toth.
\newblock {Solid-state single-photon emitters}.
\newblock {\em Nat. Phot.}, 10(10):631--641, 2016.

\bibitem{Kucsko2013}
G.~Kucsko, P.~C. Maurer, N.~Y. Yao, M.~Kubo, H.~J. Noh, P.~K. Lo, H.~Park, and
  M.~D. Lukin.
\newblock Nanometre-scale thermometry in a living cell.
\newblock {\em Nature}, 500(7460):54--58, 2013.

\bibitem{Schirhagl2014}
R.~Schirhagl, K.~Chang, M.~Loretz, and C.~L. Degen.
\newblock Nitrogen-{{Vacancy Centers}} in {{Diamond}}: {{Nanoscale Sensors}}
  for {{Physics}} and {{Biology}}.
\newblock {\em Annu. Rev. Phys. Chem.}, 65(1):83--105, 2014.

\bibitem{Loubser1978}
J.~H.~N. Loubser and J.~A. {Van Wyk}.
\newblock {Electron spin resonance in the study of diamond}.
\newblock {\em Rep. Prog. Phys.}, 41(8):1201--1248, 1978.

\bibitem{Gruber1997}
A.~Gruber, A.~Dr{\"{a}}benstedt, C.~Tietz, L.~Fleury, J.~Wrachtrup, and C.~{von
  Borczyskowski}.
\newblock {Scanning Confocal Optical Microscopy and Magnetic Resonance on
  Single Defect Centers}.
\newblock {\em Science}, 276(5321):2012--2014, 1997.

\bibitem{Jelezko2004}
F.~Jelezko, T.~Gaebel, I.~Popa, A.~Gruber, and J.~Wrachtrup.
\newblock {Observation of Coherent Oscillations in a Single Electron Spin}.
\newblock {\em Phys. Rev. Lett.}, 92(7):1--4, 2004.

\bibitem{Neumann2013}
P.~Neumann et~al.
\newblock {High-Precision Nanoscale Temperature Sensing Using Single Defects in
  Diamond}.
\newblock {\em Nano Lett.}, 13(6):2738--2742, 2013.

\bibitem{Delord2017}
T.~Delord, L.~Nicolas, M.~Bodini, and G.~H{\'e}tet.
\newblock {Diamonds levitating in a Paul trap under vacuum : Measurements of
  laser-induced heating via NV center thermometry}.
\newblock {\em Appl. Phys. Lett.}, 111(1):013101, 2017.

\bibitem{Doherty2014}
M.~W. Doherty et~al.
\newblock Electronic properties and metrology applications of the diamond
  ${\mathrm{nv}}^{\ensuremath{-}}$ center under pressure.
\newblock {\em Phys. Rev. Lett.}, 112(4):047601, 2014.

\bibitem{Dolde2011}
F.~Dolde, H.~Fedder, M.~W. Doherty, F.~Rempp, G.~Balasubramanian, F.~Reinhard,
  F.~Jelezko, and J.~Wrachtrup.
\newblock {Sensing electric fields using single diamond spins}.
\newblock {\em Nat. Phys.}, 7(6):459--463, 2011.

\bibitem{Wang2012}
Z.-H. Wang, G.~{de Lange}, D.~Rist{\'e}, R.~Hanson, and V.~V. Dobrovitski.
\newblock {Comparison of dynamical decoupling protocols for a nitrogen-vacancy
  center in diamond}.
\newblock {\em Phys. Rev. B}, 85(15):155204, 2012.

\bibitem{Taylor2008}
J.~M. Taylor, P.~Cappellaro, L.~Childress, L.~Jiang, D.~Budker, P.~R. Hemmer,
  A.~Yacoby, R.~Walsworth, and M.~D. Lukin.
\newblock {High-sensitivity diamond magnetometer with nanoscale resolution}.
\newblock {\em Nat. Phys.}, 4(10):810--816, 2008.

\bibitem{Farfurnik2016}
D.~Farfurnik, A.~Jarmola, L.~M. Pham, Z.~H. Wang, V.~V. Dobrovitski, R.~L.
  Walsworth, D.~Budker, and N.~Bar-Gill.
\newblock {Improving the coherence properties of solid-state spin ensembles via
  optimized dynamical decoupling}.
\newblock {\em Proc. SPIE}, 9900:99000N, 2016.

\bibitem{Farfurnik2018}
D.~Farfurnik, A.~Jarmola, D.~Budker, and N.~Bar-Gill.
\newblock {Spin ensemble-based AC magnetometry using concatenated dynamical
  decoupling at low temperatures}.
\newblock {\em J. Opt. (United Kingdom)}, 20(2):024008, 2018.

\bibitem{Genov2020}
G.~T. Genov, Y.~Ben-Shalom, F.~Jelezko, A.~Retzker, and N.~Bar-Gill.
\newblock Efficient and robust signal sensing by sequences of adiabatic chirped
  pulses.
\newblock {\em Phys. Rev. Res.}, 2(3):033216, aug 2020.

\bibitem{Jelezko2006}
F.~Jelezko and J.~Wrachtrup.
\newblock {Single defect centres in diamond: A review}.
\newblock {\em Physica Status Solidi A}, 203(13):3207--3225, 2006.

\bibitem{Zhang2020}
Y.~Zhang, Z.~Li, Y.~Feng, H.~Guo, H.~Wen, J.~Tang, and J.~Liu.
\newblock {High-sensitivity DC magnetic field detection with ensemble NV
  centers by pulsed quantum filtering technology}.
\newblock {\em Opt. Express}, 28(11):16191, 2020.

\bibitem{Dreau2011}
A.~Dr{\'e}au, M.~Lesik, L.~Rondin, P.~Spinicelli, O.~Arcizet, J.~F. Roch, and
  V.~Jacques.
\newblock {Avoiding power broadening in optically detected magnetic resonance
  of single NV defects for enhanced dc magnetic field sensitivity}.
\newblock {\em Phys. Rev. B}, 84(19):1--8, 2011.

\bibitem{Wolf2015}
T.~Wolf, P.~Neumann, K.~Nakamura, H.~Sumiya, T.~Ohshima, J.~Isoya, and
  J.~Wrachtrup.
\newblock {Subpicotesla diamond magnetometry}.
\newblock {\em Phys. Rev. X}, 5(4):1--10, 2015.

\bibitem{Rondin2014}
L.~Rondin, J.-P. Tetienne, T.~Hingant, J.-F. Roch, P.~Maletinsky, and
  V.~Jacques.
\newblock {Magnetometry with nitrogen-vacancy defects in diamond}.
\newblock {\em Rep. Prog. Phys.}, 77(5):056503, 2014.

\bibitem{Hong2013}
S.~Hong, M.~Grinolds, L.~Pham, D.~Sage, L.~Luan, R.~Walsworth, and A.~Yacoby.
\newblock Nanoscale magnetometry with nv centers in diamond.
\newblock {\em MRS Bulletin}, 38:155--161, 2013.

\bibitem{Barry2020}
J.~F. Barry, J.~M. Schloss, E.~Bauch, M.~J. Turner, C.~A. Hart, L.~M. Pham, and
  R.~L. Walsworth.
\newblock Sensitivity {{Optimization}} for {{NV}}-{{Diamond Magnetometry}}.
\newblock {\em Rev. Mod. Phys.}, 92(1):015004, 2020.

\bibitem{Fescenko2020}
I.~Fescenko, A.~Jarmola, I.~Savukov, P.~Kehayias, J.~Smits, J.~Damron,
  N.~Ristoff, N.~Mosavian, and V.~M. Acosta.
\newblock Diamond magnetometer enhanced by ferrite flux concentrators.
\newblock {\em Phys. Rev. Res.}, 2(2):023394, 2020.

\bibitem{Laraoui2010}
A.~Laraoui, J.~S. Hodges, and C.~A. Meriles.
\newblock Magnetometry of random ac magnetic fields using a single
  nitrogen-vacancy center.
\newblock {\em Appl. Phys. Lett.}, 97(14):143104, 2010.

\bibitem{Laraoui2011}
A.~Laraoui, J.~S. Hodges, C.~A. Ryan, and C.~A. Meriles.
\newblock Diamond nitrogen-vacancy center as a probe of random fluctuations in
  a nuclear spin ensemble.
\newblock {\em Phys. Rev. B}, 84(10):104301, 2011.

\bibitem{Barry2016}
J.~F. Barry, M.~J. Turner, J.~M. Schloss, D.~R. Glenn, Y.~Song, M.~D. Lukin,
  H.~Park, and R.~L. Walsworth.
\newblock Optical magnetic detection of single-neuron action potentials using
  quantum defects in diamond.
\newblock {\em Proc. Nat. Acad. Sci. USA}, 113(49):14133--14138, 2016.

\bibitem{Fujiwara2020}
M.~Fujiwara et~al.
\newblock Real-time nanodiamond thermometry probing in vivo thermogenic
  responses.
\newblock {\em Sci. Adv.}, 6(37):9636, 2020.

\bibitem{Poulsen21}
A.~F.~L. Poulsen, J.~D. Clement, J.~L. Webb, R.~H. Jensen, K.~Berg-S{\o}rensen,
  A.~Huck, and U.~L. Andersen.
\newblock Optimal control of a nitrogen-vacancy spin ensemble in diamond for
  sensing in the pulsed domain.
\newblock {\em arXiv:2101.10049}, 2021.

\bibitem{Ahmadi2017}
S.~Ahmadi, H.~A.~R. El-Ella, J.~O.~B. Hansen, A.~Huck, and U.~L. Andersen.
\newblock {Pump-Enhanced Continuous-Wave Magnetometry using Nitrogen-Vacancy
  Ensembles}.
\newblock {\em Phys. Rev. Applied}, 8(3):034001, 2017.

\bibitem{Ahmadi2017SI}
S.~Ahmadi, H.~A.~R. El-Ella, J.~O.~B. Hansen, A.~Huck, and U.~L. Andersen.
\newblock {Supplementary for "Pump-Enhanced Continuous-Wave Magnetometry using
  Nitrogen-Vacancy Ensembles"}.
\newblock {\em Phys. Rev. Applied}, 8(3):034001, 2017.

\bibitem{Zhu2017}
M.~Zhu, J.~Li, M.~Toda, and T.~Ono.
\newblock Microfabrication of a scanning probe with nv centers in a selectively
  grown diamond thin film through a xenon difluoride etching process.
\newblock {\em J. Micromech. Microeng.}, 27:125007, 2017.

\bibitem{Mi2020}
S.~Mi, M.~Kiss, T.~Graziosi, and N.~Quack.
\newblock Integrated photonic devices in single crystal diamond.
\newblock {\em J. Phys. Photonics}, 2(4):042001, 2020.

\bibitem{Rui2018}
Z.~Rui, Z.~Binbin, W.~Lei, G.~Hao, T.~Jun, and L.~Jun.
\newblock Optimization method of the limit sensitivity for the diamond nv color
  center magnetometer.
\newblock {\em Micronanoelectronic Technology}, 55(9):683--699, 2018.

\bibitem{Nobauer2015}
T.~N{\"o}bauer, A.~Angerer, B.~Bartels, M.~Trupke, S.~Rotter, J.~Schmiedmayer,
  F.~Mintert, and J.~Majer.
\newblock {Smooth Optimal Quantum Control for Robust Solid-State Spin
  Magnetometry}.
\newblock {\em Phys. Rev. Lett.}, 115(19):190801, 2015.

\bibitem{Rembold2020}
P.~Rembold, N.~Oshnik, M.~M. M{\"u}ller, S.~Montangero, T.~Calarco, and E.~Neu.
\newblock {Introduction to quantum optimal control for quantum sensing with
  nitrogen-vacancy centers in diamond}.
\newblock {\em AVS Quant. Sci.}, 2(2):024701, 2020.

\bibitem{Mizuno2018}
K.~Mizuno, M.~Nakajima, H.~Ishiwata, Y.~Masuyama, T.~Iwasaki, and M.~Hatano.
\newblock Wide-field diamond magnetometry with millihertz frequency resolution
  and nanotesla sensitivity.
\newblock {\em Aip Advances}, 8(12):125316, 2018.

\bibitem{McCloskey2019}
D.~J. McCloskey, N.~Dontschuk, D.~A. Broadway, A.~Nadarajah, A.~Stacey, J.-P.
  Tetienne, L.~C.~L. Hollenberg, S.~Prawer, and D.~A. Simpson.
\newblock Enhanced widefield quantum sensing with nitrogen-vacancy ensembles
  using diamond nanopillar arrays.
\newblock {\em ArXiv:1902.02464}, 2019.

\bibitem{Simpson2017}
D.~A. Simpson, E.~Morrisroe, J.~M. McCoey, A.~H. Lombard, D.~C. Mendis,
  F.~Treussart, L.~T. Hall, S.~Petrou, and L.~C.~L. Hollenberg.
\newblock Non-neurotoxic nanodiamond probes for intraneuronal temperature
  mapping.
\newblock {\em Acs Nano}, 11(12):12077--12086, 2017.

\bibitem{Shames2015}
A.~I. Shames, V.~Y. Osipov, J.~P. Boudou, A.~M. Panich, H.~J. Von~Bardeleben,
  F.~Treussart, and A.~Y. Vul'.
\newblock Magnetic resonance tracking of fluorescent nanodiamond fabrication.
\newblock {\em J. Phys. D: Appl. Phys.}, 48(15):155302, 2015.

\bibitem{Webb2021}
J.~L. Webb, L.~Troise, N.~W. hansen, L.~F. Frellsen, C.~Osterkamp, F.~Jelezko,
  S.~Jankuhn, J.~Meijer, K.~Berg-S{\o}rensen, J.-F. Perrier, A.~Huck, and U.~L.
  Andersen.
\newblock High-speed microcircuit and synthetic biosignal widefield imaging
  using nitrogen vacancies in diamond.
\newblock {\em arXiv:2107.14156}, 2021.

\bibitem{Zhang2021}
C.~Zhang, M.~Shagieva, F.~Widmann, M.~Kuebler, V.~Vorobyov, P.~Kapitanova,
  E.~Nenasheva, R.~Corkill, O.~Rohrle, K.~Nakamura, H.~Sumiya, S.~Onoda,
  J.~Isoya, and J.~Wrachtrup.
\newblock Diamond magnetometry and gradiometrty towards subpicotesla dc field
  measurement.
\newblock {\em Phys. Rev. Appl.}, 15(6):20, 2021.

\end{thebibliography}

 \end{document}